\documentclass{article}

\usepackage{amsmath}
\usepackage{amscd}
\usepackage{amsthm}
\usepackage{amssymb}
\usepackage{latexsym}
\usepackage{eufrak}
\usepackage{euscript}
\usepackage{epsfig}
\usepackage{graphics}
\usepackage{array}
\usepackage{enumerate}
\usepackage[colorlinks=true]{hyperref}
  
\theoremstyle{theorem}

\theoremstyle{corollary}

\theoremstyle{lemma}

\theoremstyle{definition}

\theoremstyle{proof}

\theoremstyle{remark}

\newcommand{\bel}[1]{\begin{equation}\label{#1}}

\newcommand{\be}{\begin{equation}}
\newcommand{\ba}{\begin{eqnarray}}
\newcommand{\ea}{\end{eqnarray}}

\newcommand{\bi}{\bibitem}
\newcommand{\qe}{\end{equation}}

\begin{document}
\title{Structural  distance and evolutionary relationship of networks}
\author{Anirban Banerjee\\[2ex]
 {\small Max Planck Institute for Molecular Genetics,}\\ 
{\small Ihnestrasse 63-73, 14195 Berlin, Germany}\\
 {\tiny banerjee@molgen.mpg.de}}
\maketitle

\begin{abstract}

Evolutionary mechanism in a self-organized system cause some functional changes that force  to adapt new conformation of the interaction pattern between the components of that system. 
Measuring the structural differences one can retrace the evolutionary relation between two systems.  
We present a method to quantify the topological distance between two networks of different sizes, finding that the architectures of the networks are more similar within the same class  than the outside of their class.
With 43 metabolic networks of different species, we show that the evolutionary relationship can be elucidated from the structural distances.

\end{abstract}

\section*{Author's Summary}

Studying the common features and universal qualities  shared by a particular class of networks in biological and other domain is one of the important aspects for evolutionary study. To measure the topological commonality, we propose a method that quantify the difference  between  two network structures of different sizes. Applying this measurement procedure  we show that the networks from the same domain have more  similarities than others. Due to the interplay between the network architecture and dynamics, biological and other networks from different areas followed by different dynamics have different structures, where  networks constructed from same evolutionary process have structural similarities. We analyze  43 metabolic networks from different species and mark the prominent separation of three groups, Bacteria, Archaea and Eukarya. That is well captured in our findings that support the other cladistic results based on gene content and ribosomal RNA sequences.Thus we show that  how evolutionary relationship can be elucidated from the structural distances  measured by our method.

\section*{Introduction}

In self-organized systems, some hidden dynamics play a role to organize the connections between the components of that system. 
Due to the interplay between the structure and dynamics, biological and other networks from different areas followed by different dynamics are expected to have different structures while networks constructed from the same evolutionary process have structural similarities.
From structural aspects, it is important to find the answer to the question of regarding the existence of a prominent difference between different types of networks, e.g., metabolic, protein-protein interaction, power grid, co-authorship or neural networks.
Studying the common features and universal qualities shared by a particular class of biological networks is one of the important aspects for evolutionary studies. In that regard, one can think about  the differences between the networks within a same class, for instance among all metabolic networks,  and also pose a question: are two metabolic networks from two different species, being  evolutionary close more similar than others?

In the last few years different notions of the graph theory have been applied and new heuristic parameters have been introduced  to analyze the network topology,  for instance degree distribution, average path length, diameter, betweenness centrality, transitivity or clustering coefficient etc.  (see \cite{Newman2003} for details). 
Those quantities, which manage to capture particular and specific properties of the graph but  not  all the qualitative aspects, are not  good representers of the  structure and hence, with those parameters it is not possible to distinguish or compare different real networks from the point of view of topology and source of formation.
 Nowadays it is a fashion to categorize networks according to their degree distribution which is the distribution of $k_n$, the number of vertices that have degree $n$. It has been observed that most of the real networks have power-law degree distribution \cite{AlbertEtAl1999,BarabasiAlbert1999,GuimeraEtAl2005,JeongEtAl2001,JeongEtAl2000,Redner1998}, thus this notion also fails to distinguish networks from different systems.  Hence focusing on particular and specific features is not enough to reveal the structural complexity in biological and other networks. 

In this article,  we propose a method to quantify the structural differences between two networks. 
We also show that the evolutionary relationships   between the  networks can be derived   from their topological similarities captured by  this quantification.
We apply this method to the metabolic networks of 43 species and show that the phylogenic evidences can be traced from the measurement of  their structural distances.\\
 
 The basic tool we use  to characterize the qualitative topological properties of a network is the normalized graph Laplacian (in short Laplacian) spectra.  Not only the global properties of the graph structure are reflected from the Laplacian spectrum, local structures produced by certain evolutionary processes, like motif joining or duplication are also well captured by the eigenvalues of this operator \cite{BanerjeeJosta,BanerjeeJost2008a,BanerjeeJostb}. 
 Distribution of the spectrum has been considered as a qualitative representation of the structure of a graph \cite{BanerjeeJost2007}.  Comparative studies on real networks are difficult because of their  complicated,   irregular structure and different sizes.
 For any graph, all eigenvalues of the graph Laplacian operator are bounded within a specific range (0 to 2). This creates the advantage to compare the spectral plots of the graphs with different sizes. 
  Spectral plots that can distinguish the networks from different origins have been used to classify the real networks from different sources\cite{BanerjeeJost2008b}. 
 Since networks constructed from the same evolutionary process produce very similar spectral plots, the distance between spectral distributions can be considered as a measurement  of the structural differences. So it can be used to study  the evolutionary relation between the networks. Here, we  quantify  this distance with the help of an existing divergence measure (Jensen-Shannon divergence) between two distributions,  what we  consider as the  quantitative distance measure of those two structures.

  \section*{Spectrum of graph Laplacian}
  The normalized graph Laplacian (henceforth simply called the Laplacian) operator ($\Delta$) has been introduced on an undirected and unweighted graph $\Gamma$, representing a network with a vertex set $V=\{i:i=1,\dots ,N \}$. For functions $v: V \to \mathbb{R}$,  graph Laplacian\footnote{This operator has the spectrum like the  operator investigated in \cite{Chung} but it  has a different spectrum than the operator  $Lv(i):=n_i v(i)-\sum_{j, j \sim i}v(j)$ usually   studied in the graph theoretical literature as the (algebraic) graph Laplacian (see \cite{Mohar1991} for this operator).}   \cite{BanerjeeJost2008a,J2,JJ1} has been defined as
\bel{1}\Delta v(i):=  v(i) -\frac{1}{n_i} \sum_{j, j \sim i}v(j) .\qe
 A nonzero solution $u$ of the equation $\Delta u-\lambda u=0$ is called an eigenfunction for the eigenvalue $\lambda$. $\Delta$ has $N$ eigenvalues, some of them may occur with higher multiplicity. The eigenvalues of this operator are real and non-negative (because $\Delta$ is selfadjoint with respect to the product $(u,v):=\sum_{i}n_iu(i)v(i)$ and  $(\Delta u,u)\ge 0$). The smallest eigenvalue $\lambda_0=0$ %\footnote{ For the convention eigenvalues have been  ordered in nondecreasing way: $0=\lambda_0\le \lambda_1\le\lambda_2\le\dots\le\lambda_{N-1}\le2$.} 
 always, since $\Delta u=0$, for any constant function $u$ and the multiplicity of this eigenvalue is equal to the number of components with the graph. The highest eigenvalue $\lambda_{N-1}$ is bounded above i.~e.~$\lambda_{N-1}\le 2$, the equality holds {\it iff} the graph is bipertite\footnote{The distance of $\lambda_{N-1}$ from 2 reflects how the graph is far from the bipertiteness.}. Another property of the spectra of a bipartite graph is if $\lambda$ is an eigenvalue, $2-\lambda$ is also an eigenvalue of that graph, hence the spectral plot will be symmetric about 1. The first nontrivial eigenvalue ($\lambda_1$ for connected graph) tells us how easily one graph can be cut into two different  components. For the complete connected graph all nontrivial eigenvalues will be equal to $\frac{N}{N-1}$.

Along with capturing the global topological characteristics  of a network, Laplacian spectrum can reveal the  local structural properties. It also has the potential to describe different evolutionary mechanisms of graph formation. 
  For instance, a single vertex $i_0\in\Gamma$ (the simplest motif)  duplication produces eigenvalue $1$, which can be found with a very high multiplicity in many biological networks, with an eigenfunction $u_1$ that takes nonzero values at $i_0$ and its duplicate $j_0$ with $u(i_0)=1$, $u(j_0)=-1$, and vanishes at other vertices. Duplication of an edge (the motif of size two) connecting the vertices $i_1$ and $i_2$ generates the eigenvalues $\lambda_\pm=1\pm\frac{1}{\sqrt{n_{i_1}n_{i_2}}}$, and the duplication of a chain $(i_1-i_2-i_3)$  of length $3$ produces the eigenvalues $\lambda=1,1\pm\sqrt{\frac{1}{n_{i_2}}(\frac{1}{n_{i_1}}+\frac{1}{n_{i_3}})}$. The duplication of these two motifs  creates the eigenvalues which are close to 1 and symmetric about 1. For certain degrees of  the vertices the duplication of these motifs  can generate the specific eigenvalues $1\pm0.5$ and $1\pm\sqrt{0.5}$ which are also mostly observed in the spectrum of real networks. If we join a motif $\Sigma$, which has an eigenvalue $\lambda$ with an eigenfunction $u_\lambda$ that vanishes at a vertex $i\in\Sigma$, via identifying the vertex $i$ with any vertex of a graph $\Gamma$, the new graph will also have the same eigenvalue $\lambda$ with an eigenfunction that  takes the same values as $u_\lambda$ on $\Sigma$ and vanishes on other vertices. As an example, if we join   a triangle that itself has an eigenvalue 1.5  to any graph, it contributes the same eigenvalue to the new graph produced by the joining process (for more details see \cite{BanerjeeJosta,BanerjeeJost2008a,BanerjeeJostb, BanerjeeJost2007, BanerjeeJostc}).

\section*{Jensen-Shannon divergence as a measure for the structural distance}

In discrete system, Kullback-Leibler divergence measure (KL) is defined on   two probability distributions $p_1$ and $p_2$  of a discrete random variable $X$ 
%(which is a measurable function from the sample space $\Omega$), from a set $S_{+}^1(\Omega)$ of  probability distributions, where $\Omega$ is a set with some $\sigma$-algebra
as
\bel{kl}
KL(p_1,p_2)=\sum_{x\in X}p_1(x)\log\frac{p_1(x)}{p_2(x)}
\qe
Note that    Kullback-Leibler (in short K-L) divergence  measure     is not defined when $p_2=0$ and $p_1\ne 0$ for any $x\in X$. K-L divergence   is not symmetric i.e. $KL(p_1,p_2)\ne KL(p_2,p_1)$ and does not satisfy the triangle inequality, hence can not be considered as a \textit{metric}.

 Jensen-Shannon divergence measure (JS) is defined on   two probability distributions $p_1$ and $p_2$ as
 \bel{js}
 JS(p_1,p_2)=\frac{1}{2}KL(p_1,p)+\frac{1}{2}KL(p_2,p); \text{ where }p=\frac{1}{2}(p_1+p_2)
 \qe
Whereas Jensen-Shannon (in short J-S) divergence  is symmetric and unlike the K-L divergence measure, it does not have any problem to be defined when one of the probability measure is zero for some value of $x$ where the other is not (for more details see \cite{Lin1991}). Square root  of J-S divergence is a metric (for details \cite{OsterreicherVajda2003}).\\

Here we have defined the structural distance $D(\Gamma_1,\Gamma_2)$ between two different graphs $\Gamma_1$ and $\Gamma_2$,  with the  spectral distribution (of graph Laplacian) $f_1$ and $f_2$ respectively, in terms of the J-S divergence measure between  $f_1$ and $f_2$:
\bel{spec-dist}
D(\Gamma_1,\Gamma_2)=\sqrt{JS(f_1,f_2)}
\qe
Theoretically there exist isospectral graphs but they are relatively rare in real networks and qualitatively quite similar in most respects. For example, all complete bipartite graphs, $K_{m,n}$ (with $m + n=$ constant), have the same spectrum. In this case distance between those two structure will be the same. This is one drawback of this measurement.

\begin{figure}[!]
\begin{center}
\includegraphics[scale=.33]{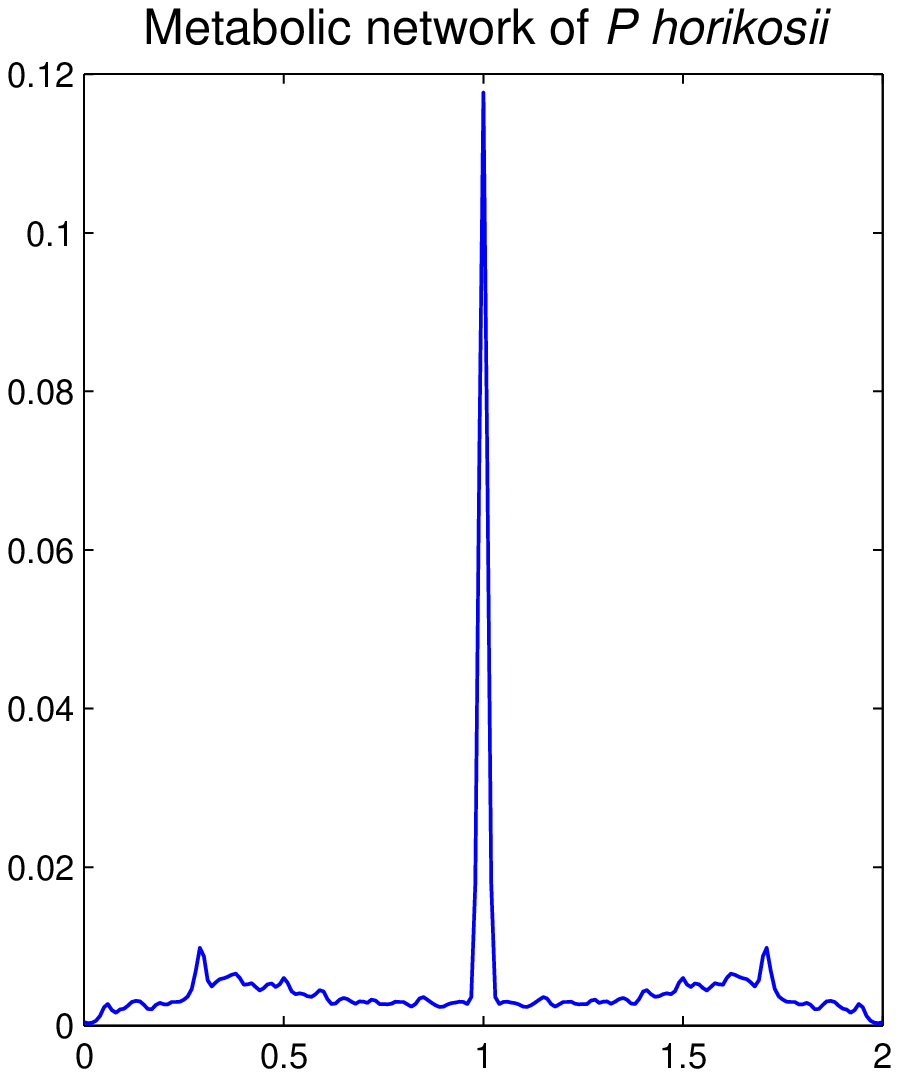}\includegraphics[scale=.33]{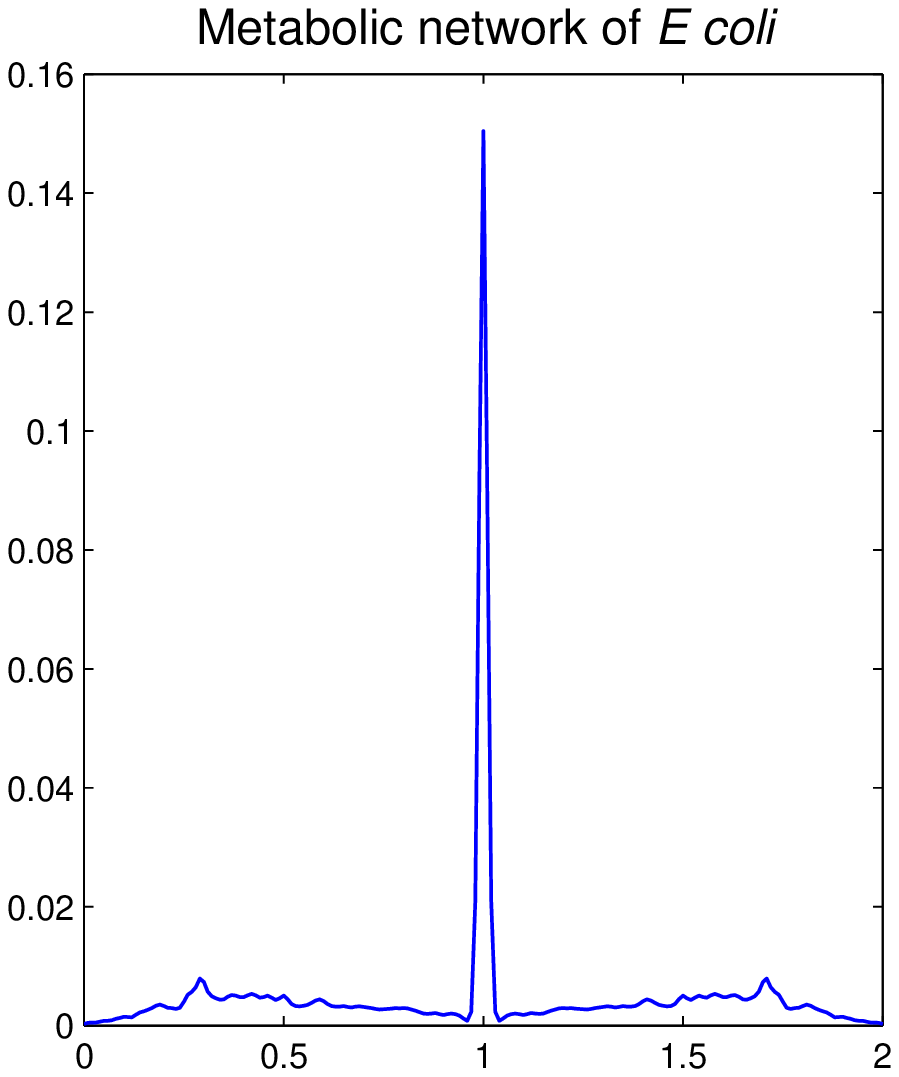}\includegraphics[scale=.33]{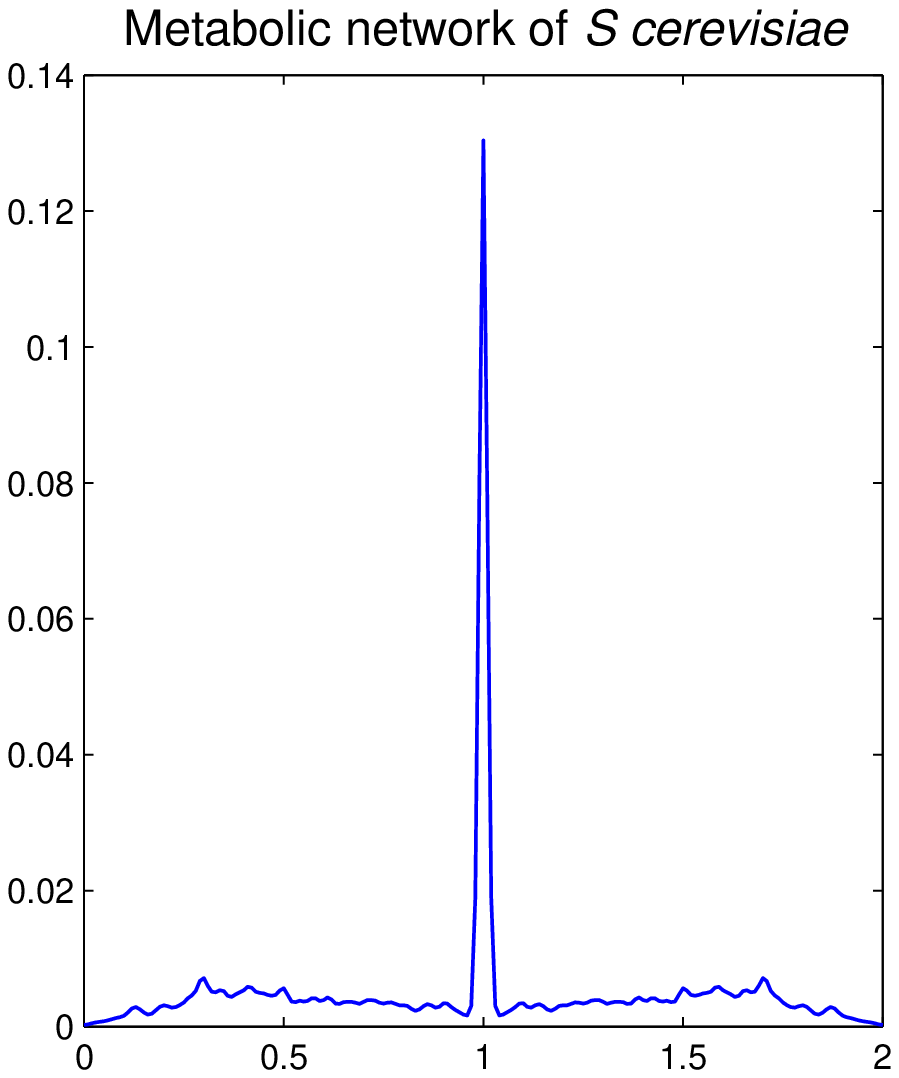}\\
\vspace{-3.2cm}
\hspace{1.5cm}(a)\hspace{3cm}(b)\hspace{3cm}(c)\\
\vspace{2.5cm}
\includegraphics[scale=.33]{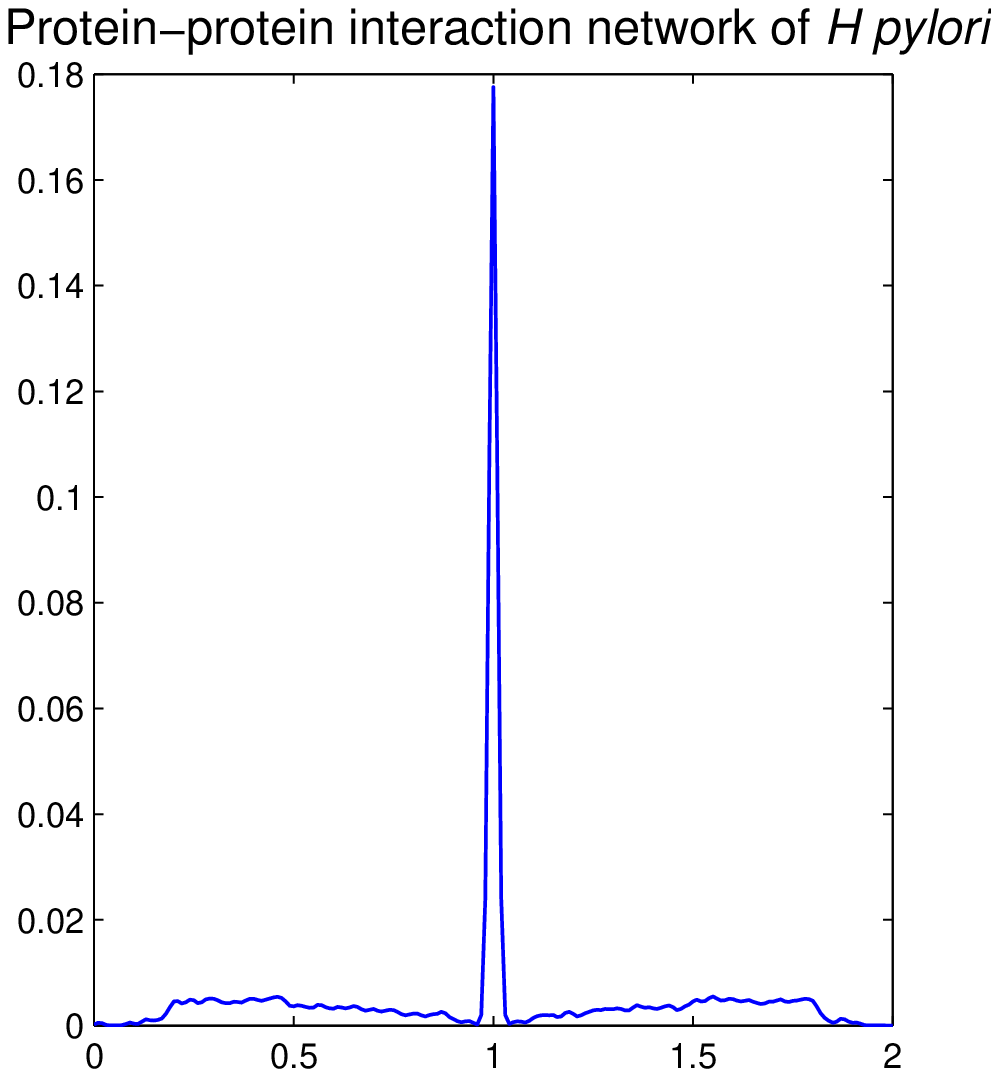}\includegraphics[scale=.33]{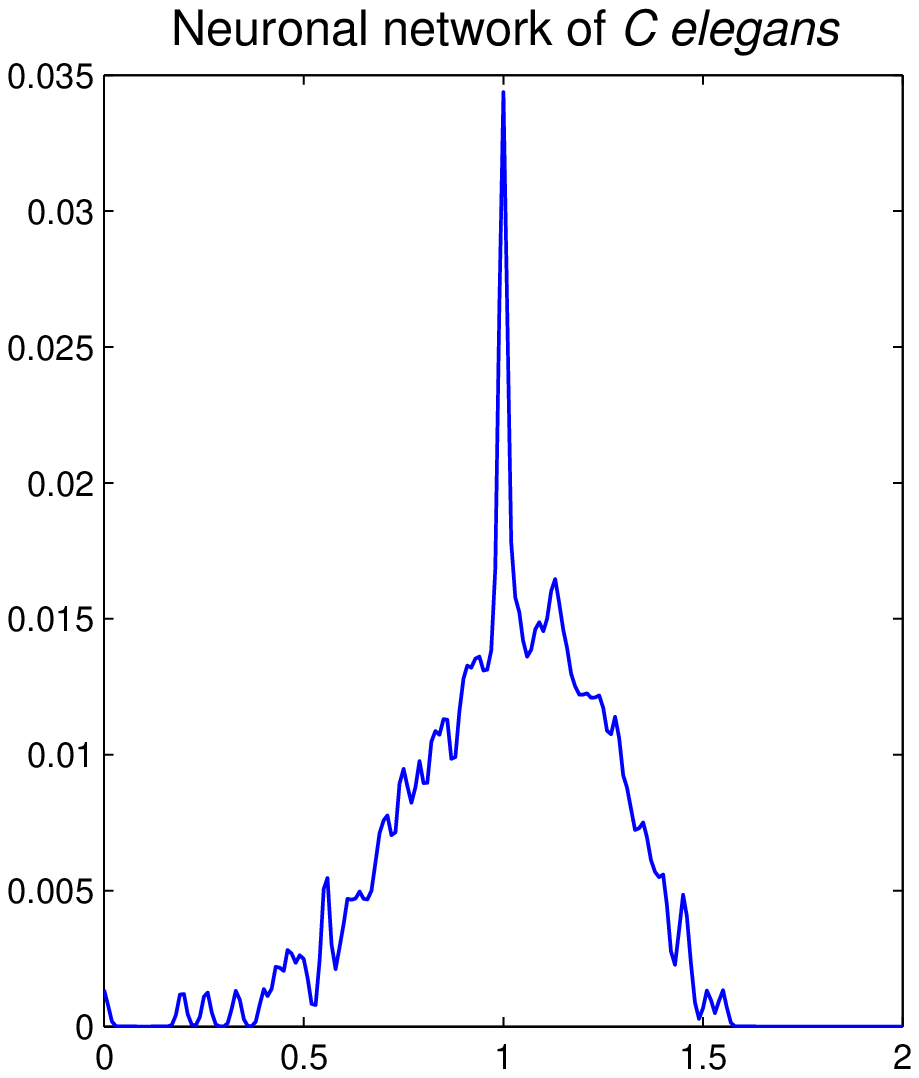}\includegraphics[scale=.33]{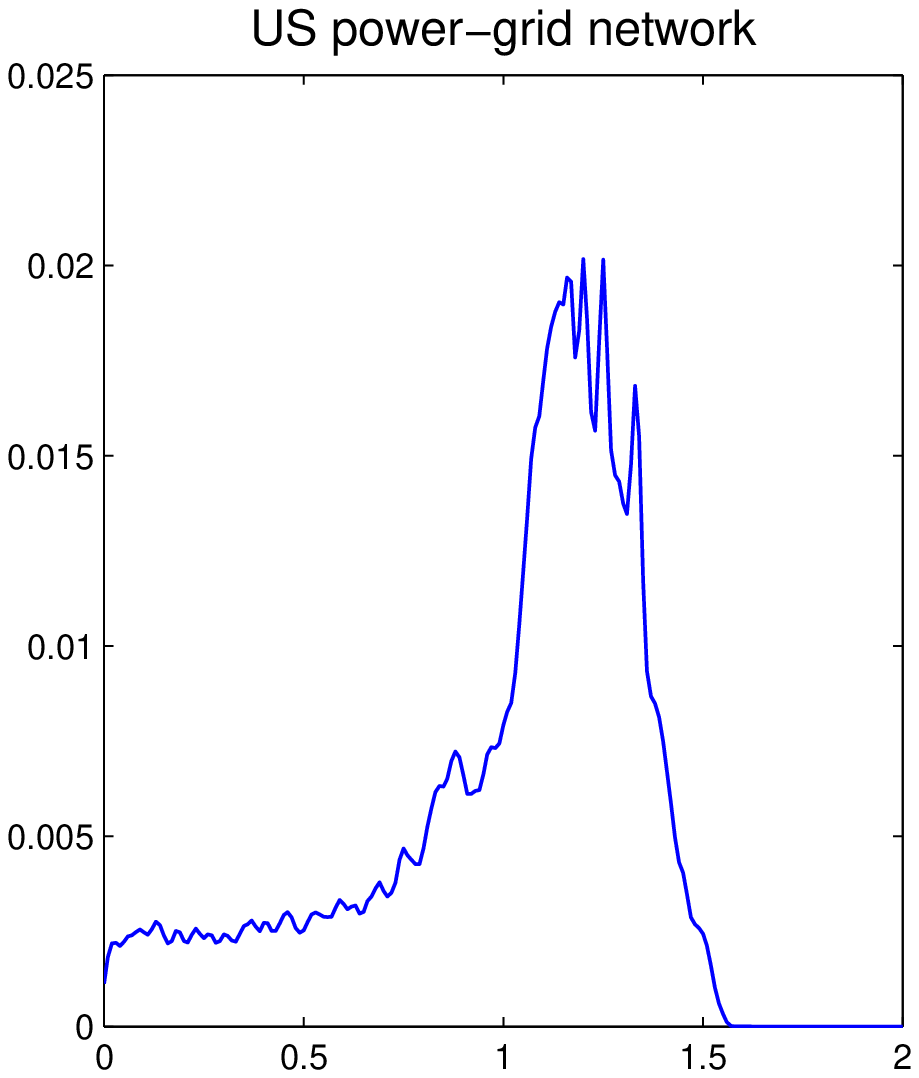}\\
\vspace{-3.2cm}
\hspace{1.5cm}(e)\hspace{3cm}(f)\hspace{3cm}(g)\\
\vspace{2.2cm}
\end{center}
\caption{ Spectral plots of the metabolic networks of (a) {\it P horikoshii}, (b) {\it E coli}, (c) {\it S cerevisiae}. The sizes of the networks are 945, 2859 and 1812 respectively.   Here the nodes represent substrates, enzymes and intermediate complexes. (d) Protein-protein interaction network of {\it H pylori}.   Network size = 710. (e) Neuronal connectivity of {\it C elegans}.  Size of the network = 297. (f)  Topology of the Western States power-grid of the United States. Network size = 4941. Here we plot the spectrum as the collection of the eigenvalues $\lambda_i$ by convolving with a Gaussian kernel (with $\sigma=0.01$). i.e.~we plot $f(x)=\sum_{\lambda_i}\frac{1}{0.01\sqrt{2\pi}}\exp(-\frac{|x-\lambda_i|^2}{0.0002})$ along the vertical axis.}
\label{specplot}
\end{figure}

\section*{Results}
Recalling the spectral similarities between different networks, metabolic networks are very similar to each other, and in comparison with the  other networks, they are closer with the protein-protein interaction networks than the neuronal or US power-grid networks in the spectral terms  \cite{BanerjeeJost2008b}. 
Due to similar mechanisms (many metabolites or proteins have the same neighbors ) of the network formation  it is expected that the metabolic networks will have similar architecture with the protein-protein interaction networks rather  than  neuronal or power-grid networks. This phenomenon is particularly reflecting in the  spectral plots (Fig.\ref{specplot}) of the metabolic networks of  {\it P horikoshii,  E coli, S cerevisiae}  with network sizes 945,  2859 and 1812 respectively,  protein-protein interaction network of {\it H pylori} with size 710, neuronal connectivity of {\it C elegans} with network size  297 and US power-grid network of size 4941 (for further reference we denote these networks by $\Gamma_{Ph}, \Gamma_{Ec}, \Gamma_{Sc},\Gamma_{Hp},\Gamma_{Ce}$ and $\Gamma_{PG}$ respectively).
Now we measure the structural distances between those networks with our metric $D$. 
The differences and similarities between those networks are clearly captured by this measurement (see the Table \ref{DistTable}). Note that each network has a  different size, but nevertheless we can measure the structural distance by comparing their spectral distributions.

 All the distances between  these three metabolic networks are closer to each other than the protein-protein interaction network, but far from the neuronal and power-grid network. It is the same for the protein-protein interaction network. The relative distance between neuronal and power-grid networks, comparative to the other networks, is less but not as close as the one between the protein-protein interaction network and metabolic networks. %On the other hand, the distances within the biological networks are  less than the distances with regard to the power-grid network. 
 These results show that we can consider our suggested metric as a suitable measure for  structural differences. 

\begin{center}
\begin{table}[ht] 
\centering 
\begin{tabular}[t] { |l ||*{6}{c}|} 
\hline 
Network &$\Gamma_{Ph}$ &$\Gamma_{Ec}$ &$\Gamma_{Sc}$ &$\Gamma_{Hp}$ &$\Gamma_{Ce}$ & $\Gamma_{PG}$ \\
\hline \hline 
$\Gamma_{Ph}$ & 0.0000 & 0.0904 &   0.0661 & 0.1694  & 0.4704  & 0.4704 \\
$\Gamma_{Ec}$ & 0.0904 & 0.0000 &  0.0641 & 0.1036  & 0.4902  & 0.5074 \\
$\Gamma_{Sc}$ & 0.0661 & 0.0641 &  0.0000 & 0.1340  & 0.4574  & 0.4738 \\ 
$\Gamma_{Hp}$ & 0.1694 & 0.1036 &  0.1340 & 0.0000  & 0.5086  & 0.5380 \\ 
$\Gamma_{Ce}$  & 0.4704 & 0.4902 &  0.4574 & 0.5086  & 0.0000  & 0.2429 \\
$\Gamma_{PG}$ & 0.4780 & 0.5074 &  0.4738 & 0.5380  & 0.2429  & 0.0000 \\
\hline 
\end{tabular} 
\caption{Distance  table between  metabolic networks of {\it P horikoshii} ($\Gamma_{Ph}$),  {\it  E coli} ($\Gamma_{Ec}$),  {\it S cerevisiae} ($\Gamma_{Sc}$); protein-protein interaction network of {\it H pylori} ($\Gamma_{Hp}$);  neuronal connectivity network of {\it C elegans} ($\Gamma_{Ce}$) and US power-grid network ($\Gamma_{PG}$). All the distances are computed using the metric $D(\Gamma_1, \Gamma_2)$.}
\label{DistTable}  
\end{table}
\end{center}

\subsection*{Evolutionary relationship from the distance measure}

\begin{figure}[!]
\begin{center}
\includegraphics[width=.5\textwidth]{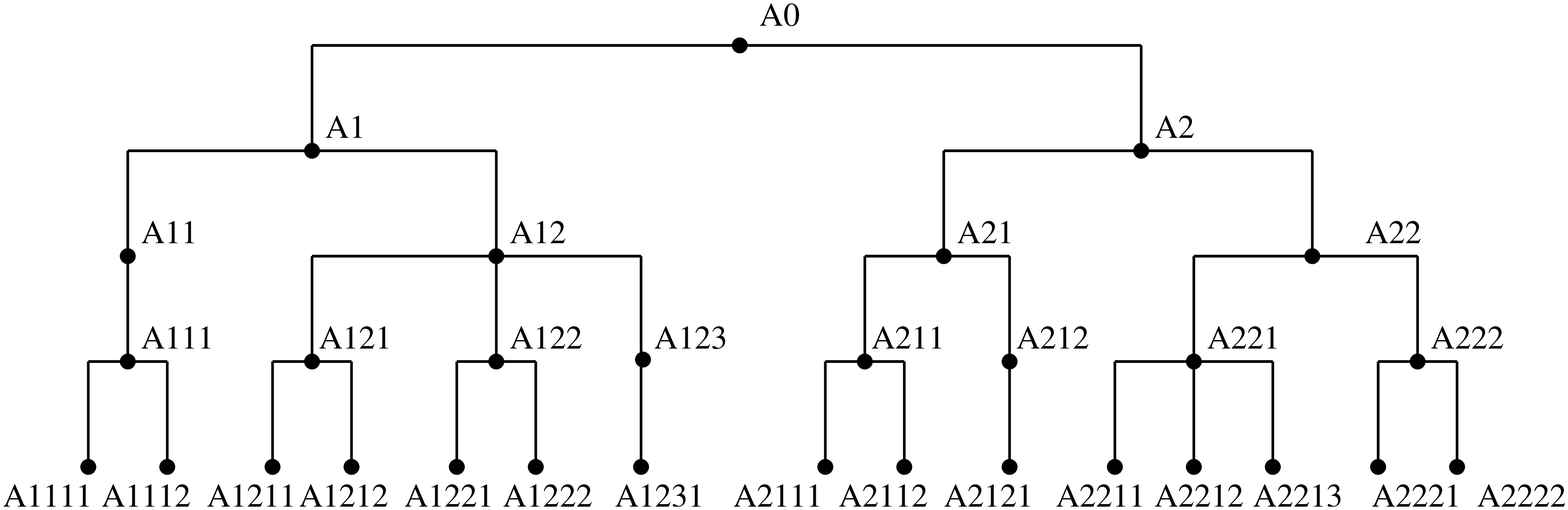}\\
\vspace{-1.2cm} 
\hspace{8cm}(a)\\
\vspace{1.2cm}
\includegraphics[width=.65\textwidth]{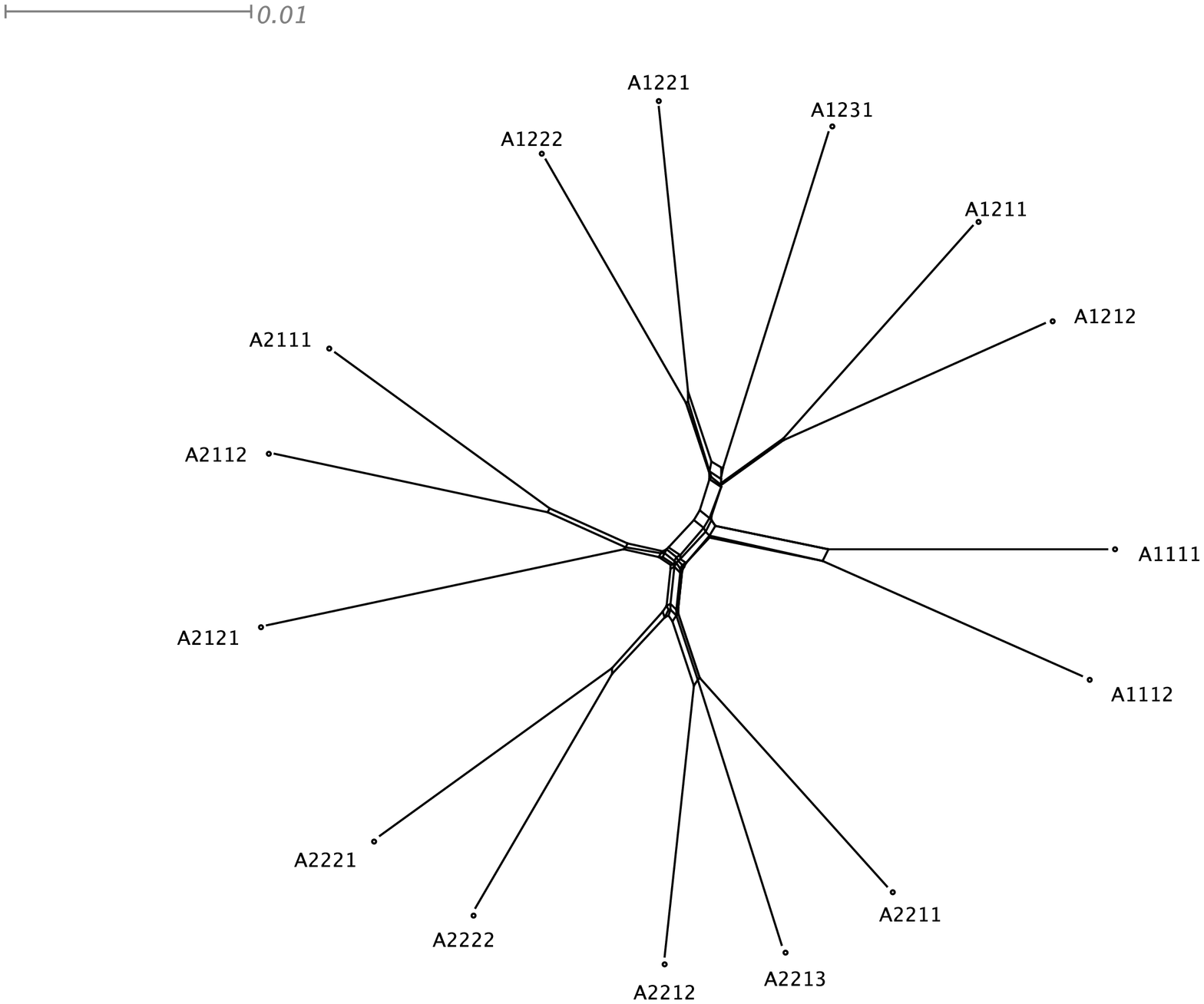}\\
\vspace{-5cm} 
\hspace{8cm}(b)\\
\vspace{5cm}
\vspace{-1.5cm}
\end{center}
\caption{(a) Evolution of a graph A0 along a definite tree:  A1 and A2 have been produced independently in the 2nd generation with a certain evolutionary process from A0. In the same way, A11 and A12 have been produced from A1 and A21, A22 from A2 and so on. Continuing in the same fashion, we end up with the graphs A1111,$\dots$,A2222 in the 5th generation. (b) The splits network for the structural distances (calculating by our proposed metric) of the graphs from the 5th generation. Each band of parallel edges indicate a split. For example, two lines represent the split \{A1111, A1112\} versus the other graphs. This tree-like splits network shows that the evolutionary relationships among those graphs is clearly captured by our distance measure. The figure has been produced by using Neighbor-Net \cite{BryantMoulton2004}. }
\label{Graph-Evo}
\end{figure}

Networks constructed from the same evolutionary process are structurally close to each other. Thus, the architectures of the networks that share the same evolutionary path are expected to be more similar than others. So to a large extent, one can elucidate the evolutionary relationships between the networks within the same system  from their structural distances. To verify this conviction we evolve a graph along a tree (see Fig.~\ref{Graph-Evo}(a)) and predict the evolutionary relations among the graphs of a generation. Here we choose the initial graph A0, a scale-free network constructed by the Barab\'{a}si--Albert's model \cite{BarabasiAlbert1999} ($m_0 = 5 \mbox{ and } m = 3 $). After a certain number of edge-rewiring, while keeping the degree of  each node the same, we produce a graph of the next generation.  Note that here all the graphs have not only the same degree distribution but also the same degree sequence. One can also choose any other evolutionary mechanism. But that would not make any significant difference in the result. We take all the graphs having been produced in the same  generation (here we choose generation 5) and  estimate the structural  distances between them using our measure $D$  (in \ref{spec-dist}). 
Now for these distances we produce a splits network \cite{Huson1998}, which can extract phylogenetic signals that are missed in other tree-representation . 
This tree-like network (see Fig.~\ref{Graph-Evo}(b)) shows  that the distances contain a prominent phylogenetic signal and clearly demonstrates the evolutionary relationships between those graphs.

%

%\begin{figure}[!]
%\begin{center}
%\includegraphics[width=.7\textwidth]{GraphEvo1.eps}\\
%\vspace{-3.2cm} 
%\hspace{8cm}(a)\\
%\vspace{3.2cm}
%\includegraphics[width=.9\textwidth]{GraphEvo2.eps}\\
%\vspace{-8cm} 
%\hspace{8cm}(b)\\
%\vspace{8cm}
%\vspace{-1.5cm}
%\end{center}
%\caption{(a) Evolution of a graph A0 along a definite tree:  A1 and A2 have been produced independently in the 2nd generation with a certain evolutionary process from A0. In the same way, A11 and A12 have been produced from A1 and A21, A22 from A2 and so on. Continuing in the same fashion, we end up with the graphs A1111,$\dots$,A2222 in the 5th generation. (b) The un-rooted tree of the graphs from the 5th generation, constructed with their structural distances (calculating by our proposed metric) using Neighbor-Net \cite{BryantMoulton2004}. This tree shows that the evolutionary relationships among those graphs is clearly captured by our distance measure.}
%\label{Graph-Evo}
%\end{figure}

\subsection*{Comparison with the other structural difference measures}

Other methods can also be used to quantify the structural similarities of the networks. A common way to compare two graph structures is to collate the independent heuristic parameters defined on them.  For this purpose, we choose the following parameters: transitivity, diameter, radius, average path length, average edge-betweeness centrality, and average node-betweeness centrality for this purpose. Now we construct a vector $V^{para}_\Gamma$, using the values of the parameters mentioned above from a graph $\Gamma$ as the components and compute the structural difference $D^{para}$ between two graphs $\Gamma_1$ and $\Gamma_2$ as
\bel{para-dist}
D^{para}(\Gamma_1,\Gamma_2) =  \parallel V^{para}_{\Gamma_1} - V^{para}_{\Gamma_2}\parallel
\qe

The other measure $D^{motif}$,  we consider, is based on the normalized Z score \cite{MiloEtAl2002} of the motif of size 3 and 4. 
It has been shown that the networks can be categorized in different superfamily \cite{MiloEtAl2004} based on the characteristic distribution of the relative frequency of their motifs.
In the similar way, we construct a vector $V^{motif}_{\Gamma}$ from a graph $\Gamma$ with the values of the normalized $Z$ score  of the motif of size 3 and 4 as the components  and compute the structural difference between two graphs $\Gamma_1$ and $\Gamma_2$ as
\bel{motif-dist}
D^{motif}(\Gamma_1,\Gamma_2) =  \parallel V^{motif}_{\Gamma_1} - V^{motif}_{\Gamma_2}\parallel
\qe

%Now, we compare the efficiency of the measure $D$ with $D^{motif}$ and $D^{para}$ to predict the evolutionary relationships among the graphs.
%With 15 realizations of the graph evolution along the tree (Fig.~\ref{Graph-Evo}), we stochastically generate 15 sets of  graphs produced in the 5th generation. Using the measures $D$, $D^{motif}$, and $D^{para}$ on the graphs in each set $I$, we construct the structural distance matrices $M_I$, $M^{motif}_I$ and $M^{para}_I$ respectively. 
%Now we apply neighbor-joining method to construct a tree from a structural distance matrix and compute
% the \textit{symmetric difference}, defined by Robinson-Foulds  \cite{RobinsonFoulds1981}, (in short R-F distance) between that tree and the true tree shown in Fig.~\ref{Graph-Evo}(a). 
%The  R-F distances  of the trees,  constructed from the cumulatively summed matrices, $\sum_{I=1}^k M_I$, $\sum_{I=1}^k  M^{motif}_I$ and $\sum_{I=1}^k M^{para}_I$ (where $k=1,\dots ,15$), from the true tree show that all of the measures are competent (Fig.~\ref{R-F}(a)).
%But the frequency distributions of the R-F distances in Fig.~\ref{R-F}(b) clearly deomnstrate that the measure $D$ is more efficient than the other two. 
%Evidently, the spectral distribution captures more qualitative properties of a network than the heuristic parametric values and the expression of the small motifs do. 

\begin{figure}[!]
\begin{center}
\includegraphics[width=.5\textwidth]{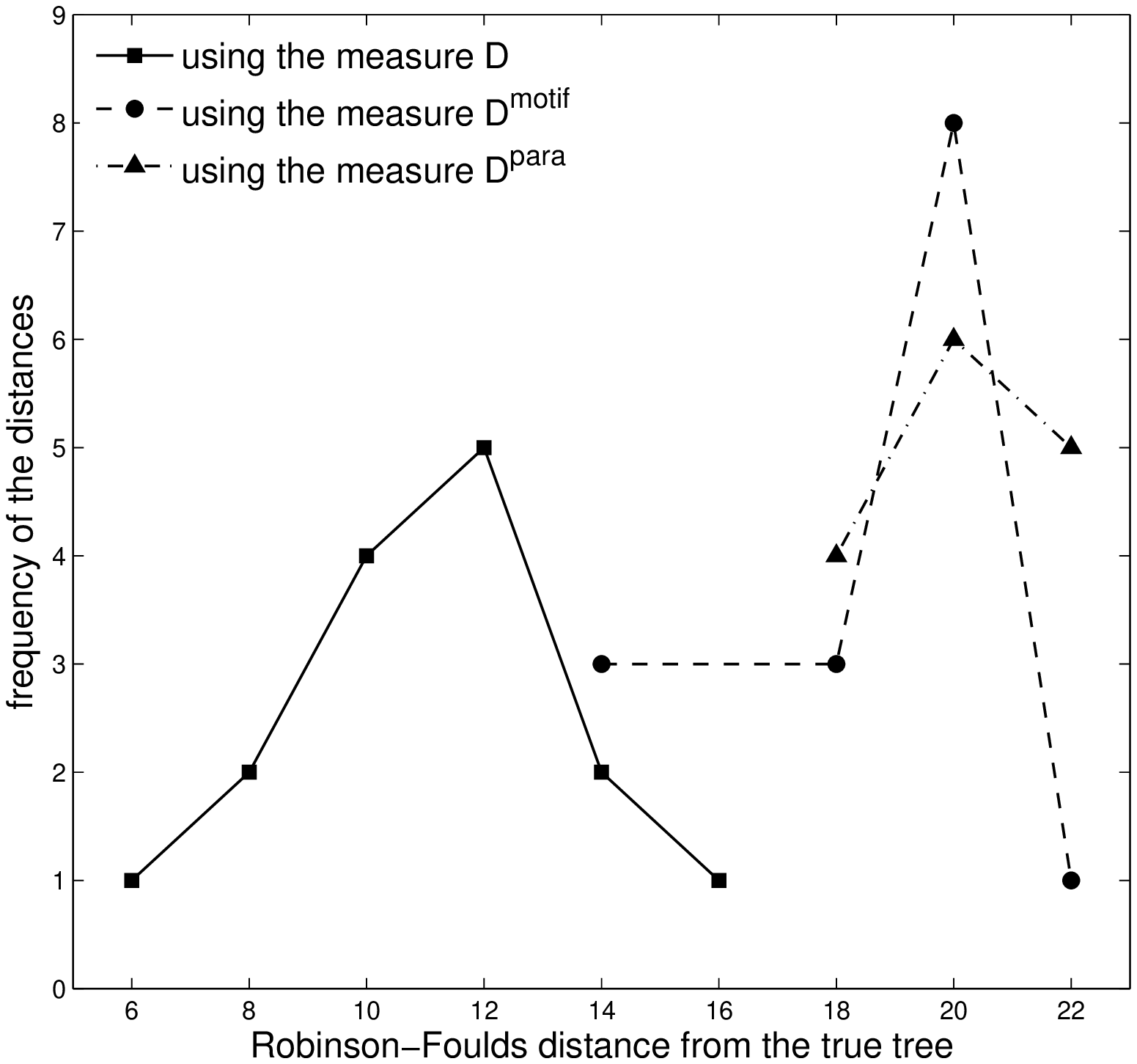}\hspace{-.5cm}
\includegraphics[width=.5\textwidth]{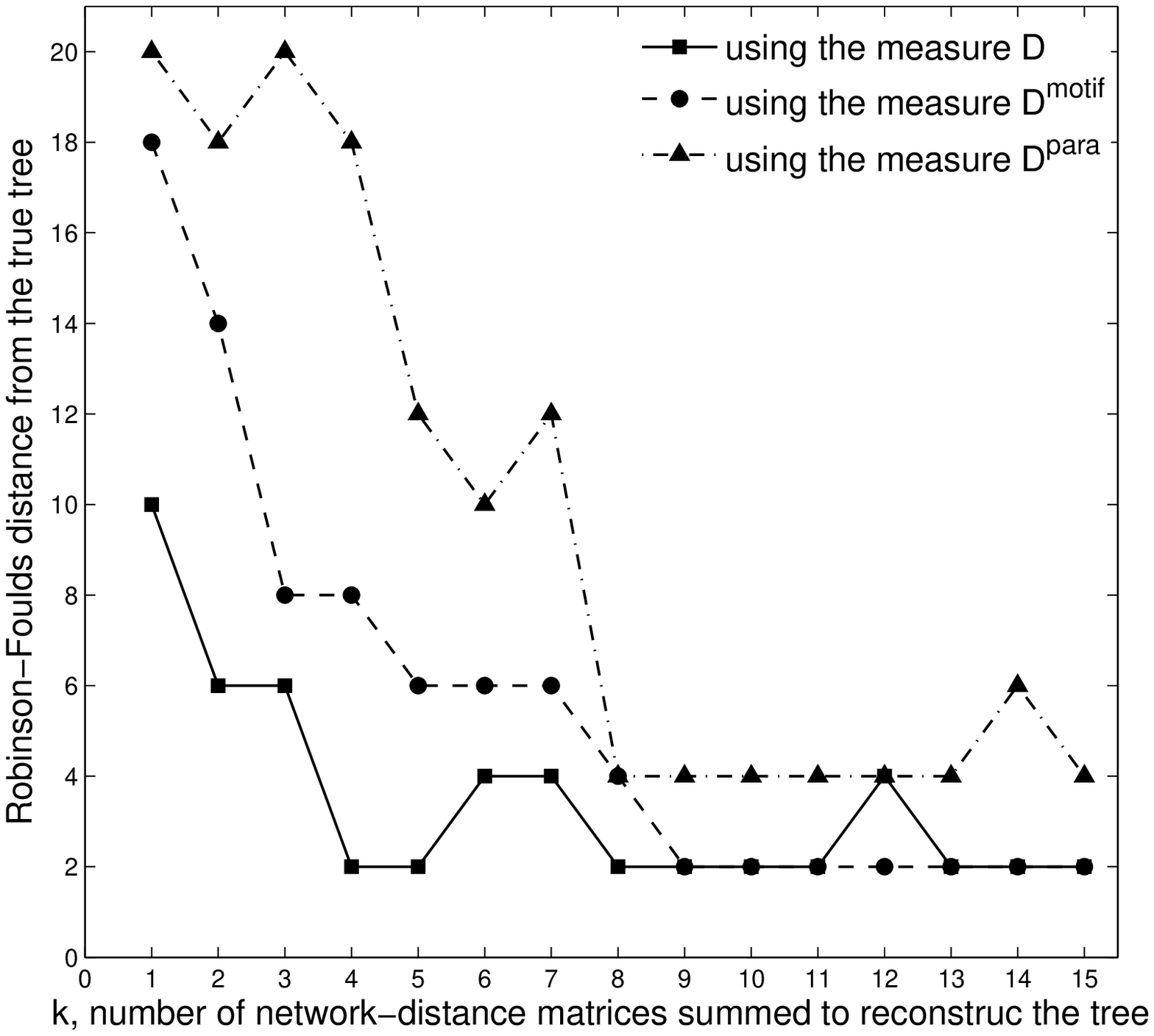}\\
\vspace{-3.8cm}
\hspace{0cm}(a)\hspace{8cm}(b)\\
\vspace{3.8cm}
\vspace{-1cm}
\end{center}
\caption{The measure $D$ is more  accurate than $D^{motif}$ and $D^{para}$. 
(a) Frequency distributions of the Robinson-Foulds distances of the trees that are constructed from graph structural-distances using $D$, $D^{motif}$, $D^{para}$  from the true tree (in Fig.~\ref{Graph-Evo}(a)). 
(b) Here we plot the Robinson-Foulds (R-F) distances along the vertical axis.  
We produce the graph distance matrices using $D$, $D^{motif}$, $D^{para}$ for every $k$ realization of graph evolution. Then we sum all the $k$ distance matrices for each measures and compute the R-F distances of the trees reconstructed from these summed matrices from the true tree.}
\label{R-F}
\end{figure}

Now we compare the efficiency of the measure $D$ with $D^{motif}$ and $D^{para}$ to predict the evolutionary relationships among the graphs.
Like previous way we compute the matrix with the distances estimated by a particular measure mentioned above between the graphs that are produced in the 5th generation of the graph evolution along the tree (Fig.~\ref{Graph-Evo}(a)). We use \textit{symmetric difference}, defined by Robinson-Foulds  \cite{RobinsonFoulds1981}, (in short R-F distance) between the tree constructed from a distance matrix using neighbor-joining method and the true tree shown in Fig.~\ref{Graph-Evo}(a). The R-F distance between two trees is the number of bipartitions that can be found in one tree but not in other one. Since our true tree contains two internal nodes (A12 and A221) of degree 4, the neighbor joining (in short N-J) tree with all the internal nodes have degree 3 always has two bipartitions which are never present in the true tree. A N-J tree that resembles the true tree most will have a R-F distance of 2 to the true tree.
Fig.~\ref{R-F}(a), which shows three frequency distributions of such R-F distances for every measures, clearly deomnstrate that the measure $D$ is more accurate than the other two.The limited accuracy can be explained by the stochasticity  in the process of graph evolution. 
In order to address whether the  accuracy is also influenced by systematic effects, we investigate the trend in the R-F distances  of the trees that are constructed using the sum of $k$ distance matrices produced by using a particular measure over $k$ realizations of graph evolution from the true tree.
The R-F distance decreases and assumes its minimum value  2 with increasing $k$ (Fig.~\ref{R-F}(b)). For this particular graph evolution, the evolutionary realationships can be perfectly recovered from the information of the $D$-measure, if the input size become large enough.
However evidently, the spectral distribution captures more qualitative properties of a network than the heuristic parametric values and the expression of the small motifs do.

%\begin{figure}[!]
%\begin{center}
%\includegraphics[width=.5\textwidth]{cumulative_distance.eps}\hspace{-.5cm}
%\includegraphics[width=.5\textwidth]{distance.eps}\\
%\vspace{-3.8cm}
%%\hspace{0cm}(a)\hspace{8cm}(b)\\
%\hspace{0cm}(a)\hspace{3cm}(b)\\
%\vspace{3.8cm}
%\vspace{-1cm}
%\end{center}
%\caption{(a) Robinson-Foulds distances of the trees, constructed from the cumulative summed matrices $\sum_I M_I$, $\sum_I M^{motif}_I$ and $\sum_I M^{para}_I$, from the true tree  in Fig.~\ref{Graph-Evo}(a). (a)Three  frequency distributions of the Robinson-Foulds distances of the trees, constructed from the matrices of the graph structural-distances, computed using the three different measures, from the true tree.}
%\label{R-F}
%\end{figure}

\subsection*{Evolutionary relationships between metabolic networks of 43 species}

\begin{figure}[!]
\begin{center}
\includegraphics[width=\textwidth]{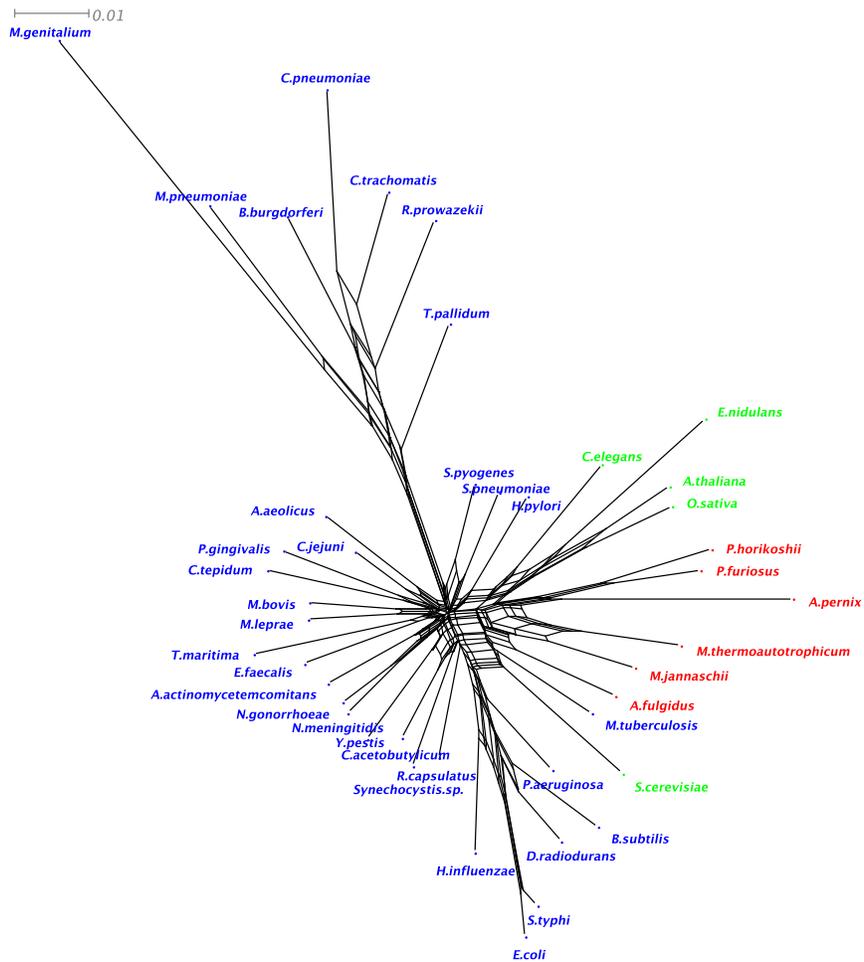}
\vspace{-1.5cm}
\end{center}
\caption{The splits network for the structural distances (calculating by the metric $D$) between the metabolic networks (of 43 species). This network shows that the distance-data is tree-like and has some phylogenetic signal.  The colors, blue, green and red indicate Bacterium, Eukaryote and Archae respectively. We use Neighbor-Net \cite{BryantMoulton2004} to produce this figure. }
\label{NNet-Phylo}
\end{figure}

\begin{figure}[!]
\begin{center}
\includegraphics[width=1\textwidth]{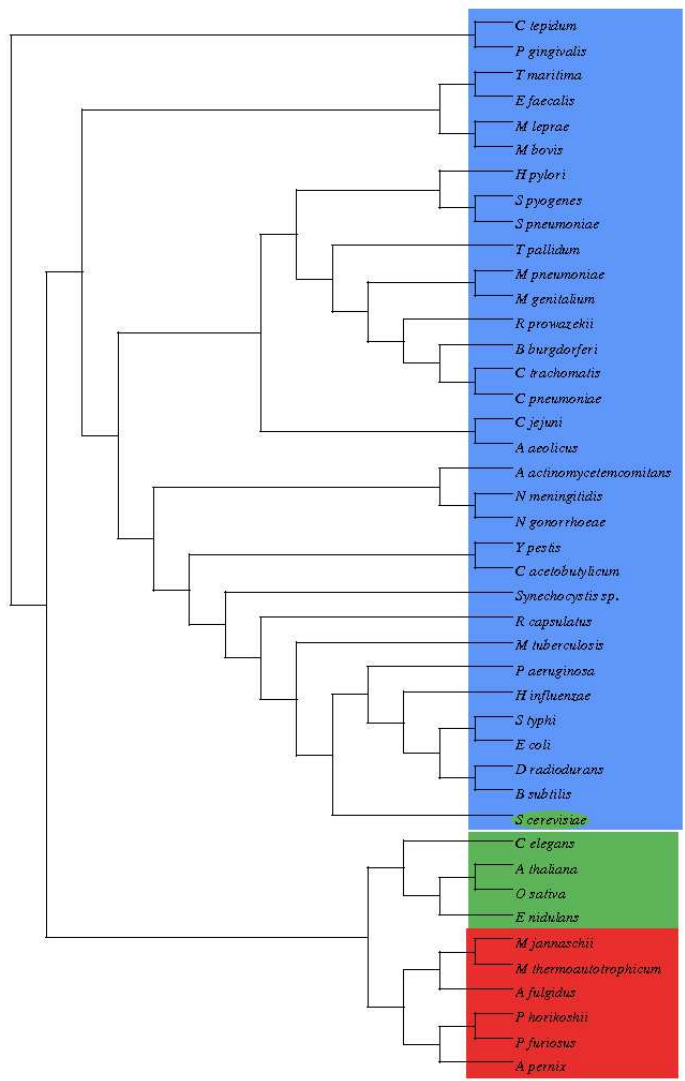}
\vspace{-1cm}
\end{center}
\caption{The un-rooted tree of metabolic networks (of 43 species) constructed with their structural distances (calculating by our proposed metric) using the neighbor-joining method. Bacterium, Eukaryote and Archae are showed by the color, blue, green and red respectively and all of them form   separate cluster within the tree. Only {\it S cerevisiae} belongs to a different group, Bacterium.}
\label{NJ-tree}
\end{figure}

Now using our structural difference measure $D$ we estimate the distances between the metabolic networks of 43 species and construct a distance matrix between them. Fig.~\ref{NNet-Phylo}, which is a splits network for these distances, supports that the data contained in that matrix has a substantial amount of phylogenetic signal and some parts of the data are tree-like. Due to the non-uniform evolutionary rate of topological change, to analyze the structural similarities among the networks of all those species we construct an unrooted tree from the mentioned distance matrix by using the neighbor-joining method. 
 This tree, which resembles highly the phylogenetic tree of those 43 species, shows different clusters according to the structural similarities of the metabolic networks (see Fig.~\ref{NJ-tree}).  
The prominent  separation of three groups, Bacteria, Archaea and Eukarya\footnote{Only Yeast  belongs to the group of Bacteria.}. That is well captured in our findings that support the other cladistic results based on gene content \cite{SnelEtAl1999} and ribosomal RNA sequences \cite{WoeseEtAl1990}. 
This is a strong evidence how evolutionary relationship is reflected from the structural similarities  which are clearly captured by the measure of the spectral distances by our metric $D$.

%\begin{figure}[!]
%\begin{center}
%\includegraphics[width=.5\textwidth]{NNet-Phylo.eps}\vspace{-0cm}
%\includegraphics[width=.49\textwidth]{Cell_Phylogram.eps}
%%\includegraphics[width=.5\textwidth]{Cell.eps}
%\vspace{-3.5cm}
%\hspace{0cm}(a)\hspace{3cm}(b)\\
%\vspace{3.5cm}
%\vspace{-0cm}
%\end{center}
%\caption{The un-rooted tree of metabolic networks (of 43 species) constructed with their structural distances (calculating by our proposed metric) using (a) Neighbor-Net \cite{BryantMoulton2004}, (b) the neighbor-joining method. The tree in (a) shows that the data of the distance matrix is tree-like and has some phylogenetic signal. Bacterium, Eukaryote and Archae have been showed by the color, blue, green and red respectively and all of them have formed separate cluster. Only {\it S cerevisiae} belongs to a different group, Bacterium.}
%\label{NNet-NJ-tree}
%\end{figure}

\subsection*{Cross validation of the tree construction against the effect of the enzyme mapping from \textit{E.~coli}}

\begin{figure}[!]
\begin{center}
\includegraphics[width=.4\textwidth]{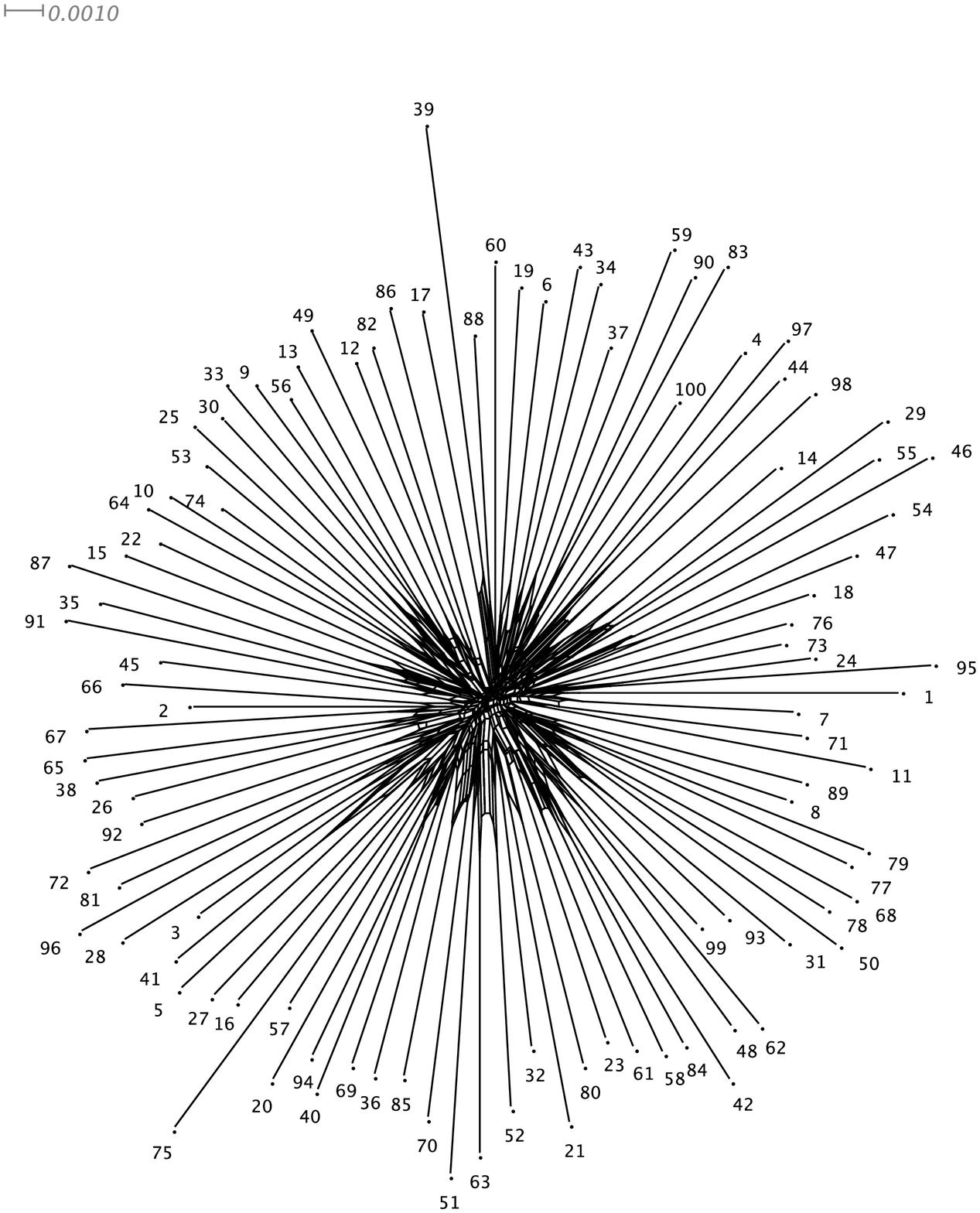}
\includegraphics[width=.59\textwidth]{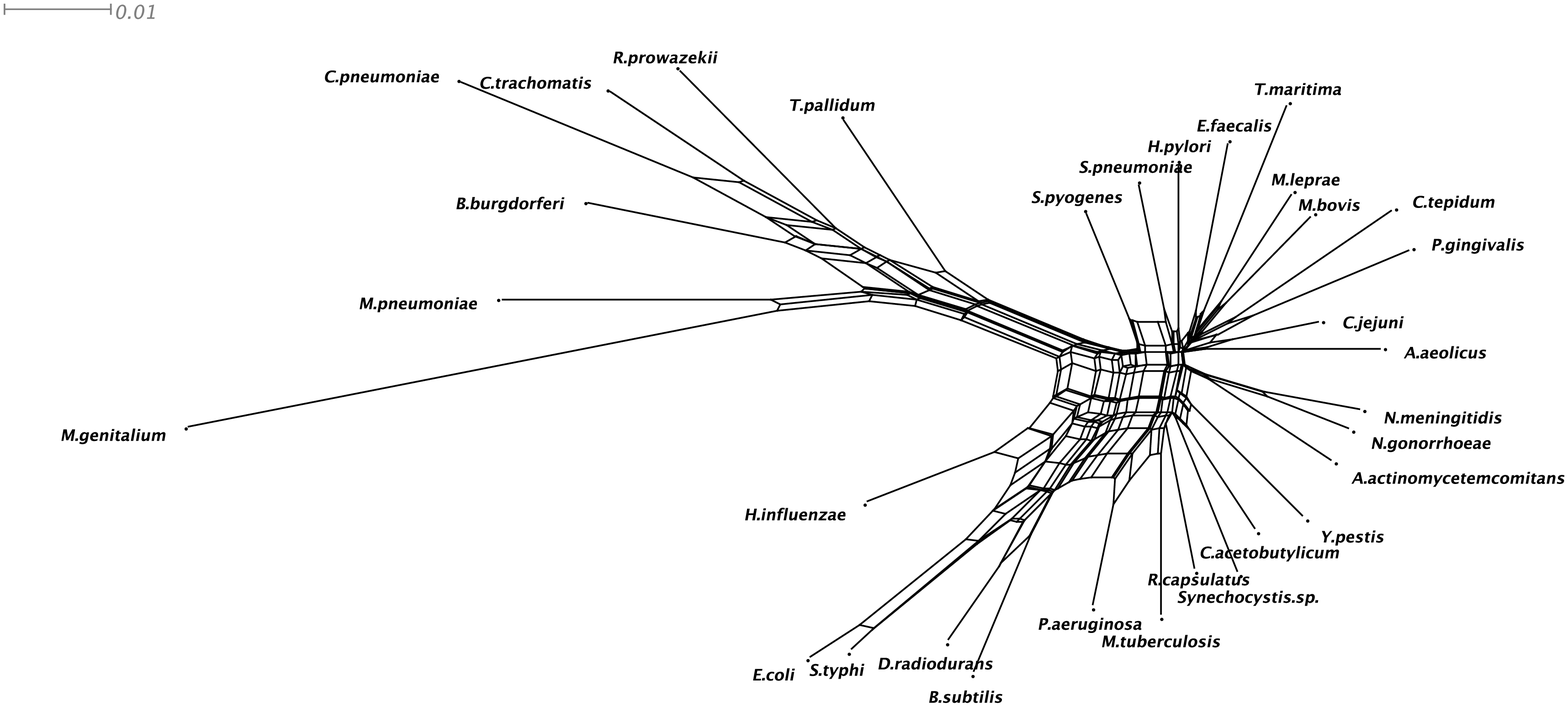}\\
\vspace{-.3cm}
\hspace{0cm}(a)\hspace{4.5cm}(b)\\
\vspace{0cm}
\vspace{-0.2cm}
\end{center}
\caption{The splits network of the structural distances between (a) 100 networks constructed by randomly deleting 5 percent of the reactions from the metabolic network of \textit{E.~coli} and (b) metabolic networks of 32 bacteria. The star-like structure of the splits network in (a), which is very different from the splits network of bacteria in (b), shows that the data of the distance matrix merely has a phylogenetic signal and the metabolic networks of bacteria are not constructed only by mapping from the \textit{E.coli}. We have used Neighbor-Net \cite{BryantMoulton2004} to construct both the splits networks.}
\label{EC-BootStrap}
\end{figure}

All the metabolic pathways in \textit{E.~coli} have been constructed independently in wetlab. But it is not always the case for the other bacteria. If an enzyme-specific gene that also exists in \textit{E.~coli} has been detected, the same metabolic reactions catalyzed by that enzyme are incorporated into the database. 
 If there are no different genes which have been reported from every other bacteria and that can make significant change in the network structure, all other metabolic networks will be very similar and  the detection of the phylogenetic relationship can be an artifact. In order to verify this fact, we reconstruct 100 networks by randomly deleting 5 percent of the reactions from the metabolic network of \textit{E.~coli} and produce a splits network of the distances between those 100 networks. The star-like structure of this splits network, which is very different from the splits network constructed from the structural distances between the metabolic networks of 32 bacteria, shows that the distances of those 100 networks merely have a phylogenetic signal (Fig.~\ref{EC-BootStrap}). Hence the evolutionary relationships can not be detected if all other metabolic networks are only mapped from the network of \textit{E.~coli}.

%
%\begin{figure}[!]
%\begin{center}
%\includegraphics[width=.6\textwidth]{EC_BStap_5_100_1.eps}\\
%\vspace{-8cm} 
%\hspace{8cm}(a)\\
%\vspace{8cm}
%\vspace{-1.5cm}\includegraphics[width=1\textwidth]{JS_WholeCell_Bact.eps}\\
%\vspace{-6cm} 
%\hspace{10cm}(b)\\
%\vspace{6cm}
%\vspace{-.5cm}
%\end{center}
%\caption{The un-rooted tree of (a) 100 networks constructed by randomly deleting 5 percent of the reactions from the metabolic network of \textit{E.~coli} and (b) metabolic networks of 32 bacteria. Both the trees have been constructed with the structural distances (calculated by our proposed method) of the networks using Neighbor-Net \cite{BryantMoulton2004}. The star-like structure of the tree in (a), which is very different from the tree of bacteria in (b), shows that the data of the distance matrix merely has a phylogenetic signal and the metabolic networks of bacteria are not constructed only by mapping from the \textit{E.coli}.}
%\label{EC-BootStrap}
%\end{figure}

\section*{Discussion}

Here we suggest a method to compare the  architecture of the networks with different sizes, an aspect  causing the main problem for the comparison.   
With a defined metric, we quantify their structural similarities based on the spectral distribution which captures the qualitative  properties of the underlying graph topology which can emerge from the evolutionary process like motif duplication or joining, random rewiring, random edge deletion etc.
In spite of the network reconstruction error (see source of the data),  this method elucidate the evolutionary relationships between the metabolic networks constructed from 43 different species.
To explore the evolutionary relationships in other domains like  language and society structure and in other biological areas, this approach can also  be used.

\section*{Methods}

\subsection*{Sources of the data}

In this article we use the data set which are freely available. We access the metabolic data (used in \cite{JeongEtAl2000}) of 43 species from  \url{http://www.nd.edu/~networks/}. At the time of database construction genomes of 25 species (18 bacteria, 2 eukaryotes and 5 archaea) had been completely sequenced while the remaining 18 species underwent this process partially. But the analysis of the errors \cite{JeongEtAl2000} suggest that there would not be a drastic change in the final result.
We use the network data for the protein-protein interaction of {\it Helicobacter pylori} from \url{http://www.cosinproject.org/} and neuronal connectivity (used in \cite{WattsStrogatz1998,WhiteEtAl1986} ) of {\it C elegans} from \url{http://cdg.columbia.edu/cdg/datasets}.

\subsection*{Network construction from the data set}
Due to incomplete sequencing of the genome of different species, many biological data are incomplete and they contain statistical errors. 
To capture a more appropriate (i.e. with less error) network architecture  we focus on the giant component. It is very   probable that this part of the network is constructed from the mostly studied metabolic pathways, hence consists more complete  data and capture most of the qualitative properties of the original complete network.  
Moreover, in our analysis we consider the underlying undirected graphs of the real networks which are directed in many  cases.  The reduced graph itself carries a lot of structural information that is quite informative about the network, but one can easily extend this method to  directed networks for having more accurate results. 

\subsection*{Compute the distribution of the spectrum}
After computing the spectrum of a network we convolve with a kernel 
$g(x,\lambda)$ and get the distribution by normalizing the function 
 \be
f(x)=\int g(x,\lambda)\sum_k \delta (\lambda, \lambda_k)d\lambda= \sum_k g(x,\lambda_k)
\qe
Here we use the Gaussian kernel $\frac{1}{\sqrt{2\pi \sigma^2}}\exp(-\frac{(x-m_x)^2}{2\sigma^2})$ with $\sigma=.01$ for all computation. Choosing other types of kernels  does not change the result significantly.\\

\subsection*{Clustering of the metabolic networks by constructing an unrooted tree}

Since we are interested only to get the clusters among all those metabolic networks according to their structural distance, an unrooted tree is our interest, thus the neighbor-joining method is adequate to choose for the construction. We calculate the $D(\Gamma_i,\Gamma_j)$ for each pair of those networks  $(\Gamma_i,\Gamma_j)$  and build  a distance matrix. We use the software package PHYLIP \cite{Felsenstein1996} and SplitsTree \cite{Huson1998} for the tree construction.
The branching distance is not important for our purpose, hence we ignore the branch length while plotting the tree.

%\subsection*{Evolve a graph along a tree}

\subsection*{Compute the normalized $Z$ score of a motif}

The normalized $Z$ score of a motif  of a network is the  normalized relative frequency of that motif, compared to its expression in the randomized version of the same network. The statistical significance of a motif $\sigma$ is presented by its $Z$ score,
\bel{Z-score}
Z_{\sigma}=\frac{N^{real}_{\sigma} - \langle N^{rand}_{\sigma} \rangle}{SD(N^{rand}_{\sigma})},
\qe
where $N^{real}_\sigma$ is the number of times the motif $\sigma$ appears in the network, and $\langle N^{rand}_\sigma\rangle$ and $SD(N^{rand}_\sigma)$ are the mean and standard deviation of its appearance in the ensemble of randomized networks. Hence the normalized $Z$ score of a motif $\sigma$ is $Z_\sigma/(\sum_\sigma Z_\sigma^2)^{1/2}$. Here, with the help of the software mfinder1.2, which is freely available on \url{http://www.weizmann.ac.il/mcb/UriAlon/}, we calculate the $Z$ score of each motif of size 3 and 4, and normalize them over all.

\section*{Acknowledgments}
 The author is thankful to Martin Vingron, Thomas Manke, Roman Brinzanik, Sitabhra Sinha, Monojit Choudhur for valuable discussions. A special thank to Hannes Luz for giving  the useful suggestions regarding phylogenetic tree construction. The author is also thankful to Antje Gl\"uck for the helpful comments on preparing the manuscript. Thanks to the VolkswagenStiftung for the funding to support this project. 

\bibliographystyle{plain}

\end{document}